 \documentclass[final,5p,times,twocolumn]{elsarticle}

\usepackage{hyperref}
\usepackage{amssymb}
\usepackage{amsmath}
\usepackage{siunitx}
\usepackage{color} 

\newcommand{\NiS}{NiS$_2$}
\newcommand{\TC}{\ensuremath{T_{\text C}}}
\newcommand{\TN}{\ensuremath{T_{\text N}}}

\newcommand{\nS}{\ensuremath{n_{\text S}}}
\newcommand{\muS}{\ensuremath{\mu_{\text S}}}
\newcommand{\meV}{\milli\electronvolt}
\newcommand{\kB}{\ensuremath{k_{\text B}}}
\newcommand{\nV}{\ensuremath{n_{\text V}}}
\newcommand{\DeltaI}{\ensuremath{\Delta_{\text I}}}
\newcommand{\DB}{\ensuremath{\Delta_{\text B}}}
\newcommand{\DS}{\ensuremath{\Delta_{\text S}}}
\newcommand{\K}{\kelvin}

\newcommand{\D}{\mathrm{d}}
\newcommand{\dd}[1]{\D{#1}}

\newcommand{\sven}[1]{%
{#1}}

\journal{Journal of Magnetism and Magnetic Materials}

\begin{document}

\begin{frontmatter}



\title{Atomic Diffusion in the Surface State of Mott Insulator \NiS}


\author{C. Clark}
\author{S. Friedemann}

\address{HH Wills Laboratory, University of Bristol, BS8 1TL Bristol, UK}

\begin{abstract}
We present resistivity measurements of Mott insulator \NiS\ single crystals after heat treatment. We find a strong increase of the low temperature resistivity that relaxes back towards the pristine behaviour over several days with a time constant of \SI{45}{\hour} at room temperature. The low temperature resistivity has previously been shown to be dominated by surface conduction \cite{Thio1994}. Consequently, the changes induced by heat treatment are attributed to changes to surface states. Our results suggest the creation of vacancies in the surface that re refilled from the bulk via atomic diffusion. We estimate a diffusion constant of $D\approx\SI{1e-10}{\meter\squared\per\second}$ at room temperature. We identify sulphur vacancies as the most likely to form via oxidisation of sulphur forming volatile SO$_2$ during heat treatment. Our results point towards these sulphur vacancies to be the source of surface state localisation in \NiS.
\end{abstract}


\end{frontmatter}


%
%
\section{Introduction}
\label{sec:Intro}
\begin{figure}%
	\begin{center}
		\includegraphics[width=.5\columnwidth]{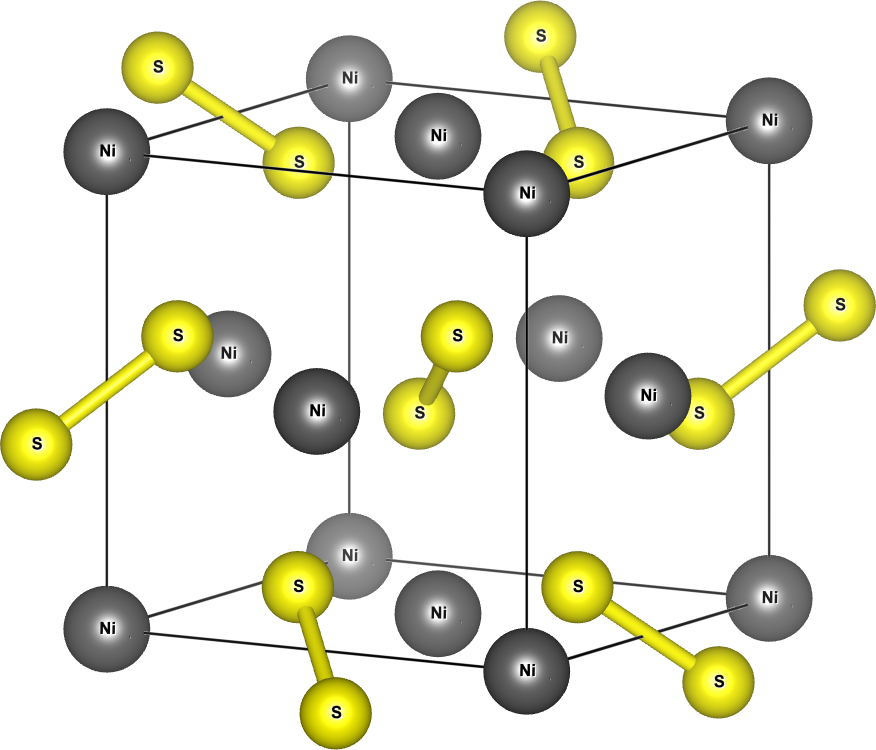}%
	\end{center}
	\caption{Crystal structure of cubic \NiS.}%
	\label{fig:NiS2}%
\end{figure}

\NiS\ is one of the best examples of a Mott-Hubbard metal-insulator system \cite{Imada1998,Wilson1985a,Honig1998}. It can be understood analogous to NiO, the canonical Mott-Hubbard insulator \cite{Mott1949}: The sulphur atoms in \NiS\ form dimers S$^{2-}_2$ with a total valence of two (cf.~Fig.~\ref{fig:NiS2}). As a consequence, Ni is divalent with a half-filled $e_g$ band as it is in NiO. It is these Ni states that are split by correlations into the lower Hubbard band $\approx \SI{1}{\electronvolt}$ below the chemical potential $\mu$ and the upper Hubbard band $\approx \SI{3.5}{\electronvolt}$ above $\mu$ \cite{Kunes2010}.
Whilst the sulphur bonding bands $p_{\sigma}$ are well below the lower Hubbard band the antibonding band $p_{\sigma}^{\star}$ is centred at $\approx \SI{1}{\electronvolt}$ above $\mu$. Thus the charge gap forms between the lower Hubbard band and the sulphur antibonding band classifying \NiS\ as a charge-transfer insulator \cite{Zaanen1985}.

Magnetic order is found at low temperatures in \NiS: Whilst \NiS\ is paramagnetic at room temperature, it orders antiferromagnetically below  around \SI{38}{\K} and eventually becomes a weak ferromagnet below \SI{30}{\K} \cite{Matsuura2003}.

The bulk insulating state in \NiS\ can be tuned by either Se substitution or pressure. This presents \NiS\ as a prototypical system to study Mott-Hubbard physics where fundamental questions can be addressed. In fact, recent quantum oscillation measurements at high pressures succeeded in demonstrating the emergence of a large Fermi surface encompassing the full electron count in the metallic high-pressure phase \cite{Friedemann2015}.

The insulating state in bulk \NiS\ at ambient pressure has a gap of $\DB \approx \SI{100}{\meV}$ as seen in optical measurements and in activated behaviour in  resistivity measurements at high temperatures \cite{Kautz1972,Kwizera1980}. This gap is derived from the above band positions by taking into account the band width of the lower Hubbard band and the sulphur antibonding band.
Surprisingly, a deviation from activated behaviour is seen below \SI{200}{\K}: the resistivity $\rho(T)$ shows a shoulder at $\approx\SI{150}{\K}$ with a plateau at $\approx\SI{100}{\K}$ and a second strong increase of $\rho(T)$ below \SI{20}{\K}.

It has been suggested that the complex temperature dependence of the resistivity originates from a surface layer with metallic behaviour above \SI{50}{\K} and insulating behaviour below \SI{50}{\K} in addition to the activated behaviour for bulk conduction \cite{Thio1994}. Hall effect measurements show a diminishing charge carrier concentration as the temperature is lowered from room temperature to \SI{100}{\K} in accordance with activated bulk behaviour. A saturation of carrier concentration  below \SI{100}{\K} is attributed to surface conduction dominating at low temperatures. Surface conduction has been underpinned by comparison of samples with different ratios of surface area to volume showing the expected shift in the respective contributions \cite{Thio1994}. The surface carriers have a rather low mobility of $\muS \approx \SI{1}{\centi\meter\squared\per\volt\per\second}$ but a large surface charge carrier concentration $\nS  \approx\SI{5e14}{\per\centi\meter\squared}$ amounting to $\approx\SI{1}{electron}$ per unit cell at the surface. This indicates $\approx 1$ broken bond per unit cell to generate the amount of surface charge carriers observed. 
The steep increase in resistivity at lowest temperatures $T<\SI{30}{\K}$ has  been attributed to Anderson localisation due to the increased effect of disorder \cite{Sarma2003}.
The precise origin of the charge carriers on the surface has not been identified previously.  

Here, we report resistivity measurements in combination with heat treatments to our sample. We observe an enhancement of the low temperature resistivity. This enhancement relaxes back towards the behaviour of the pristine sample over several days. These results point towards a sulphur depletion during the heat treatment that is replenished from the bulk through atomic diffusion. Our results suggest that sulphur atoms provide charge carriers to the conducting surface state whilst sulphur vacancies act as trapping potentials of $\approx\SI{1}{\meV}$. Furthermore, our results suggest that it is these vacancies that cause the localisation of surface states at lowest temperatures.

%
%
\section{Experimental Details}
Single crystals were grown in two steps following the procedure in \cite{Yao1994}: First, polycrystalline \NiS\ was produced via solid state reaction from stoichiometric ratios of high purity nickel and sulphur. Single crystals were then obtained from this starting material using the Tellurium flux technique with slight excess of sulphur to compensate for losses due to the high vapour pressure of sulphur. The resulting millimeter sized crystals were characterised with powder x-ray diffraction to have a lattice constant of \SI{5.6893(5)}{\angstrom}. Structure refinement of single crystal diffraction measurements confirmed full occupancy of sulphur and nickel sites in the bulk \cite{Friedemann2015}.

Resistivity measurements were performed in a $^4$He-cryostat using 4-point low-frequency lock-in technique with a constant current of \SI{100}{\nano\ampere} which was tested not to cause Joule heating. Contacts were made using silver loaded epoxy (Epo-Tek EE129-4).

Heat treatment of the sample was carried out at \SI{150}{\celsius} for \SI{2}{\hour} in a constant flow of $\approx\SI{0.5}{\milli\litre\per\second}$ of Argon gas with a purity of \SI{99.998}{\percent}. However, even this high level of purity can provide enough oxygen to oxidise the sulphur of the surface of our crystal as estimated in \ref{sec:oxygen}.

%
%
\section{Results}
\label{sec:Results}

\begin{figure}%
\includegraphics[width=\columnwidth]{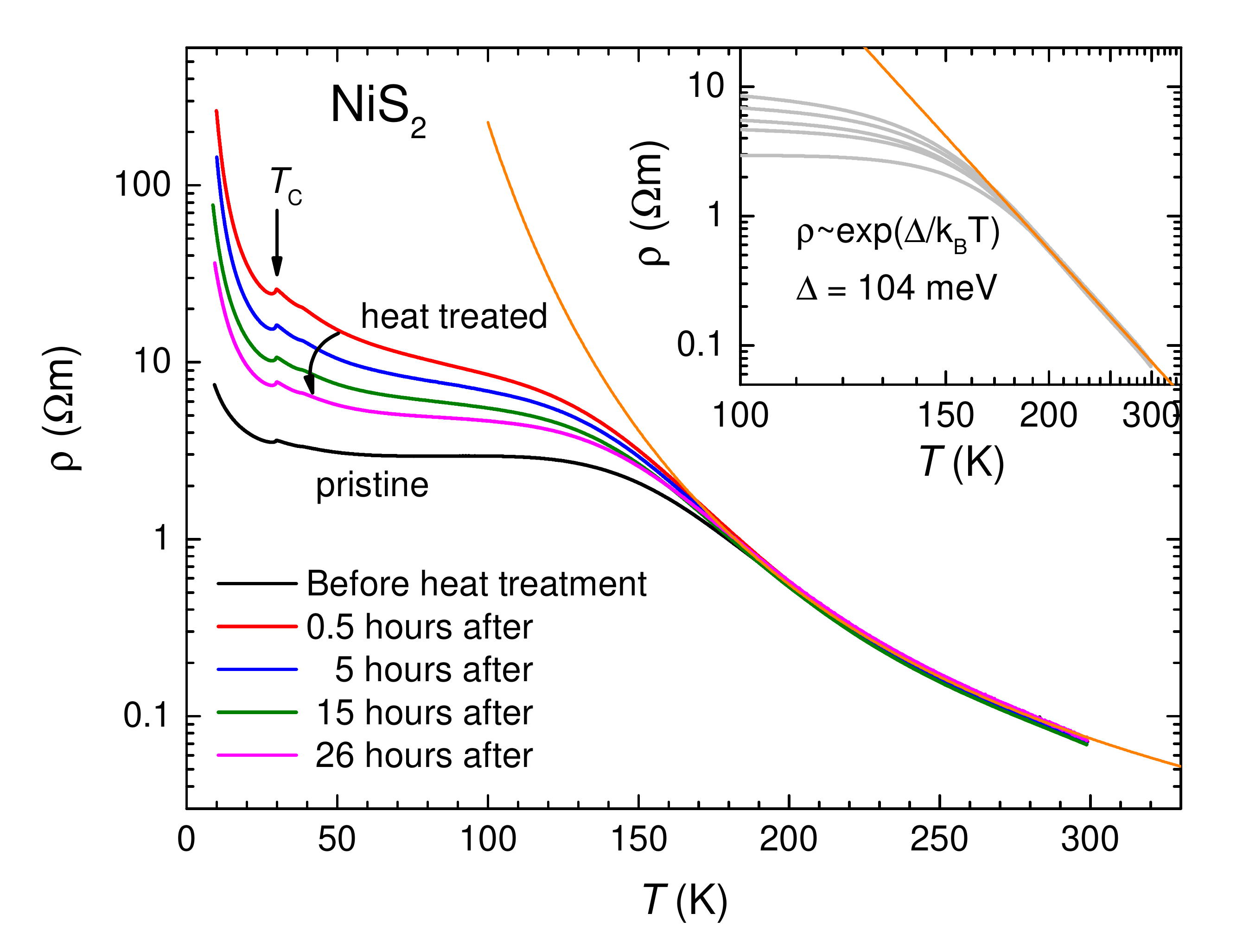}%
\caption{Temperature dependence of the resistivity of \NiS\ for our sample before and after heat treatment at the times given in the legend. These times represent time at room temperature in accordance with our model of atomic diffusion. Curved arrow indicates the time evolution after heat treatment. Inset shows same data on a logarithmic resistivity scale versus reciprocal temperature scale. Orange lines in inset and main panel represent the same fit to activated behaviour $\rho \propto \exp(\Delta/\kB T)$ above \SI{200}{\K}. Straight arrow in the main panel indicates the transition temperature of the ferromagnetic order.}%
\label{fig:RvsT}%
\end{figure}%

Our resistivity measurements of the pristine sample (black curve in Fig.~\ref{fig:RvsT}) show typical behaviour for \NiS: The resistivity increases towards lower temperatures with activated behaviour between \SIlist{200;300}{\K} and with a corresponding activation energy of $\Delta = \SI{104}{\meV}$ in good agreement with earlier reports of high-temperature behaviour \cite{Kwizera1980,Bither1968,Honig1998,Yao1996}. Deviation from the high-temperature activated behaviour is attributed to surface conduction (cf.~orange line in Fig.~\ref{fig:RvsT}) \cite{Thio1994}: The shoulder in $\rho(T)$ around \SI{150}{\K} and the plateau between \SIrange{50}{100}{\K} arises due to metallic behaviour of the surface whilst the surface conduction turns insulating at low temperatures leading to the strong increase in resistivity below \SI{50}{\K}.

\begin{figure}%
\includegraphics[width=\columnwidth]{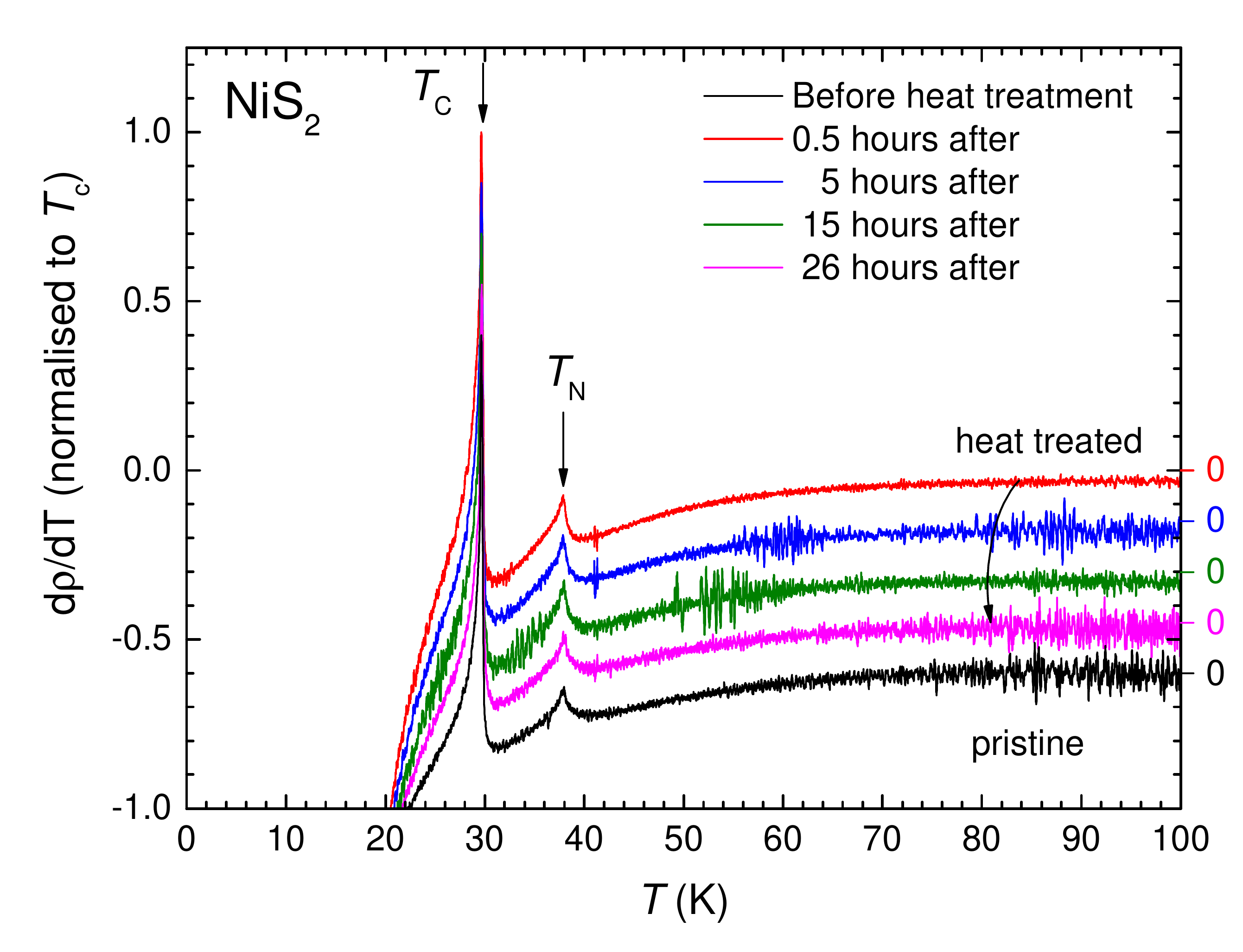}%
\caption{Derivative of the resistivity $\dd{\rho}/\dd{T}$ as a function of temperature around the magnetic phase transitions. Data after heat treatment are offset for clarity \sven{with zero for each dataset indicated on the right}. Arrows indicate peaks attributed to the onset of weak ferromagnetism at \TC\ and antiferromagnetism at \TN.}%
\label{fig:dRdT}%
\end{figure}

\sven{The surface to volume ratio of our sample is significant for the rectangular cross section of $\SI{0.5}{\milli\meter} \times \SI{0.2}{\milli\meter}$. This is in accordance with a sizeable surface contribution associated with the large deviation from the high-temperature activated behaviour below \SI{200}{\K}. For the pristine sample we observe metallic behaviour of the total resistivity $\dd{\rho}/\dd{T}>0$ around \SI{85}{\K} (cf.~Fig.~\ref{fig:dRdT}).}


A peak is observed in $\rho(T)$ at $\TC=\SI{30}{\K}$ marking the transition into the ferromagnetic state \cite{Yao1996}. 
No clear signature is seen in $\rho(T)$ for the antiferromagnetic transition. Only in the derivative $\dd{\rho}/\dd{T}$ a peak is visible at $\TN = \SI{38}{\K}$ (Fig.~\ref{fig:dRdT}) which we assign to the antiferromagnetic transition \cite{Mori1978,Thio1995,Matsuura2003}. 

We now turn to our results after heat treatment in argon for \SI{2}{\hour} at \SI{150}{\celsius}. The resistivity before and after heat treatment are compared in Fig~\ref{fig:RvsT}. Immediately after heat treatment the resistivity reveals a large enhancement of more than an order of magnitude at low temperatures whilst the high-temperature behaviour above \SI{200}{\K} remains the same. This low-temperature enhancement is found to decrease over the course of several hours to days. Such a slow relaxation is a first indication of atomic diffusion as an underlying cause. Atomic diffusion is exponentially diminished at low temperatures; thus, we count time the sample spent at room temperature between subsequent resistivity measurements presented in Fig.~\ref{fig:RvsT}. \sven{In fact, when storing the sample at low temperatures ($T\approx\SI{9}{\K}$) immediately after the heat treatment the resistivity remained constant to better than \SI{0.5}{\percent} over \SI{3}{\hour} compared to a change of almost \SI{50}{\percent} of the low temperature resistance after \SI{0.5}{\hour} storage at room temperature. This corresponds to a reduction of diffusion of more than \num{100} fold.}

The low temperature insulating behaviour shows a qualitative change after heat treatment. In general it follows activated behaviour but displays sample dependences. Many samples can be fitted with an exponential form including a power-law form of the temperature in the exponential term 
\begin{equation}
 \rho(T) = A \exp\left\{\left( \frac{\DS}{\kB T}\right)^x\right\}.
\label{eq:VRH}
\end{equation} 
Such a general form is predicted for variable range hopping \cite{Mott1968a}. For our sample the pristine resistivity can be fitted with eq.~\ref{eq:VRH} below \SI{20}{\K} where we find an exponent $x=0.8$ and a gap $\Delta_{\text S} = \SI{1}{\meV}$. This exponent deviates from the value of $x=1/2$ originally predicted  for 2D systems \cite{Mott1968a}. 

\begin{figure}%
\includegraphics[width=\columnwidth]{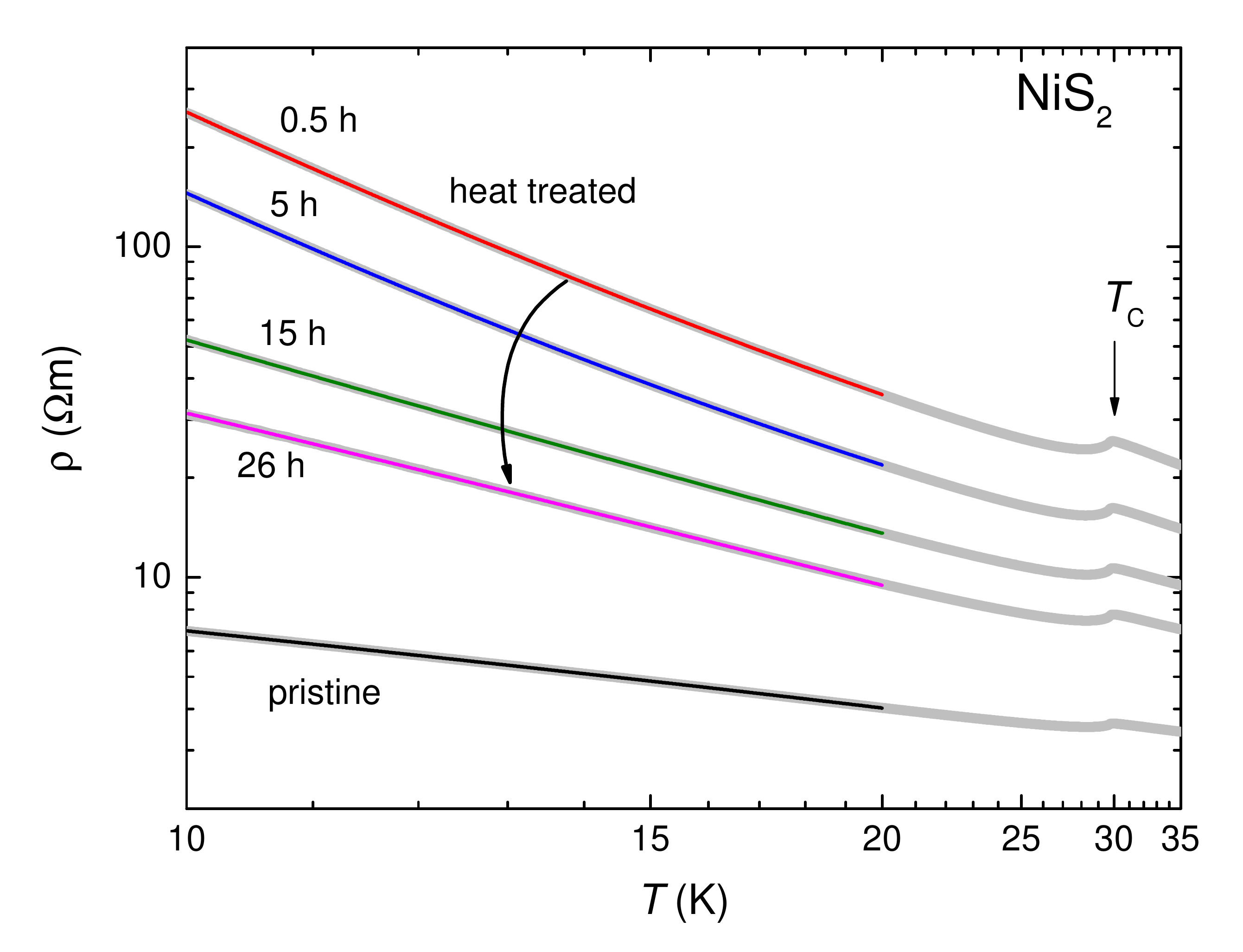}%
\caption{Low temperature behaviour of the resistivity. Coloured lines represent fits of eq.~\ref{eq:VRH} to the data between \SIlist{10;20}{\K}. Data are plotted with the resistivity on a logarithmic scale and the temperature on a reciprocal scale. On this scale an exponent $x=1$ yields a straight line whilst $x<1$ and $x>1$ manifest as downward and upward curvature, respectively.}%
\label{fig:VRH}%
\end{figure}

Heat treatment is found to alter the low temperature behaviour of the resistivity. Whilst $\rho(T)$ can still be fitted using (eq.~\ref{eq:VRH}) the exponent becomes larger and does exceed $x>1$ as can be seen from the upwards curvature towards lower temperatures in Fig.~\ref{fig:VRH} compared to the slight downward curvature of the pristine sample. 

Strong Coulomb repulsion is expected to enhance the exponent in the variable range hopping model in 2D beyond $x=1/2$ but is expected to be limited to $x<1$ \cite{Efros1975,Pollak1972,Hamilton1972}. Thus the exponent found for the pristine sample can be understood within the correlation corrections of the variable range hopping model but the behaviour after heat treatment is incompatible. This deviation remains elusive.

Given the complex behaviour of the resistivity with both bulk and surface contribution, and the surface behaviour only partially understood, we aim to analyse the time dependence of the resistivity without restricting to a specific form $\rho(T)$. We take a Drude model as the starting point with the conductivity 
\begin{equation} 
	\sigma = \frac{n e^2}{m \nu}.
\label{eq:Drude}
\end{equation} 
Here, both the charge carrier concentration $n$ and the scattering rate $\nu$ may be temperature dependent for instance with activated behaviour $n\propto\exp(\Delta/\kB T)$ and power law temperature dependence $\nu \propto T^p$, respectively while $e$ and $m$ denote the electron charge and mass, respectively. 

The enhancement of the resistivity after heat treatment at low temperatures roughly following activated behaviour (eq.~\ref{eq:VRH}) suggests a reduction of the charge carrier concentration to be dominant.
Two main scenarios are considered which are analysed with the same procedure:

i) The reduction of surface charge carrier concentration can arise from trapping of available states in impurity potentials of magnitude \DeltaI\ around vacancies created by heat treatment. This leads to a reduced surface charge carrier concentration 
\begin{equation}
	n= \left[n_0 - \nV \exp \left(-\frac{\DeltaI}{\kB T} \right) \exp \left(-\frac{t}{\tau}\right)\right]\exp\left(-\frac{\DS}{\kB T}\right)
\label{eq:nI}
\end{equation}
where the vacancy concentration \nV\ decays over time $t$ with a relaxation time $\tau$. The surface charge carrier concentration is subject to activated behaviour with a gap \DS\ as derived above leading to a combined gap of the impurity and surface conduction 
\begin{equation}
	n=n_0 \exp\left(-\frac{\DS}{\kB T}\right) - \nV \exp\left(-\frac{t}{\tau}\right) \exp\left(-\frac{\DeltaI +\DS}{\kB T}\right).
\label{eq:nIR}
\end{equation}


ii) Alternatively, the charge carrier concentration can be directly reduced. Surface states in \NiS\ are suggested to arise from dangling bonds of individual atoms. Heat treatment removing these atoms can lead to a reduction of surface charge carrier concentration 
\begin{equation}
	n=\left[n_0 - \nV \exp\left(-\frac{t}{\tau}\right)\right]\exp\left(-\frac{\DS}{\kB T}\right)
\label{eq:nS}
\end{equation}
where the surface charge carrier concentration is subject to the surface activated behaviour with gap \DS. 
Eq.~\ref{eq:nS} can be rearranged to a similar form to eq.~\ref{eq:nI} 
\begin{equation}
	n=n_0 \exp\left(-\frac{\DS}{\kB T}\right) - \nV \exp\left(-\frac{t}{\tau}\right) \exp\left(-\frac{\DS}{\kB T}\right).
\label{eq:nSR}
\end{equation}

Substituting the time dependent charge carrier concentration of either (eq.~\ref{eq:nI}) or eq.~\ref{eq:nS} into the Drude conductivity we obtain 
\begin{equation}
	\left( \sigma - \sigma_0 \right)= - \frac{e^2 \nV}{m \nu}\left[ \exp\left(-\frac{\Delta}{\kB T}\right) \exp\left(-\frac{t}{\tau}\right)\right],
\label{eq:s0s}
\end{equation}
where the gap is $\Delta = \DeltaI+\DS$ or $\Delta=\DS$ for scenario i) or ii), respectively. In both cases the initial conductivity $\sigma_0$ is associated with the initial charge carrier concentration $n_0$ before heat treatment. This allows us to analyse the data for the two scenarios at the same time where we will identify indications for scenario ii). Additional (bulk) contributions to the conductivity eq.~\ref{eq:s0s} can be absorbed into $\sigma_0$ in a two-band model of surface and bulk conductivity.

\begin{figure}%
\includegraphics[width=\columnwidth]{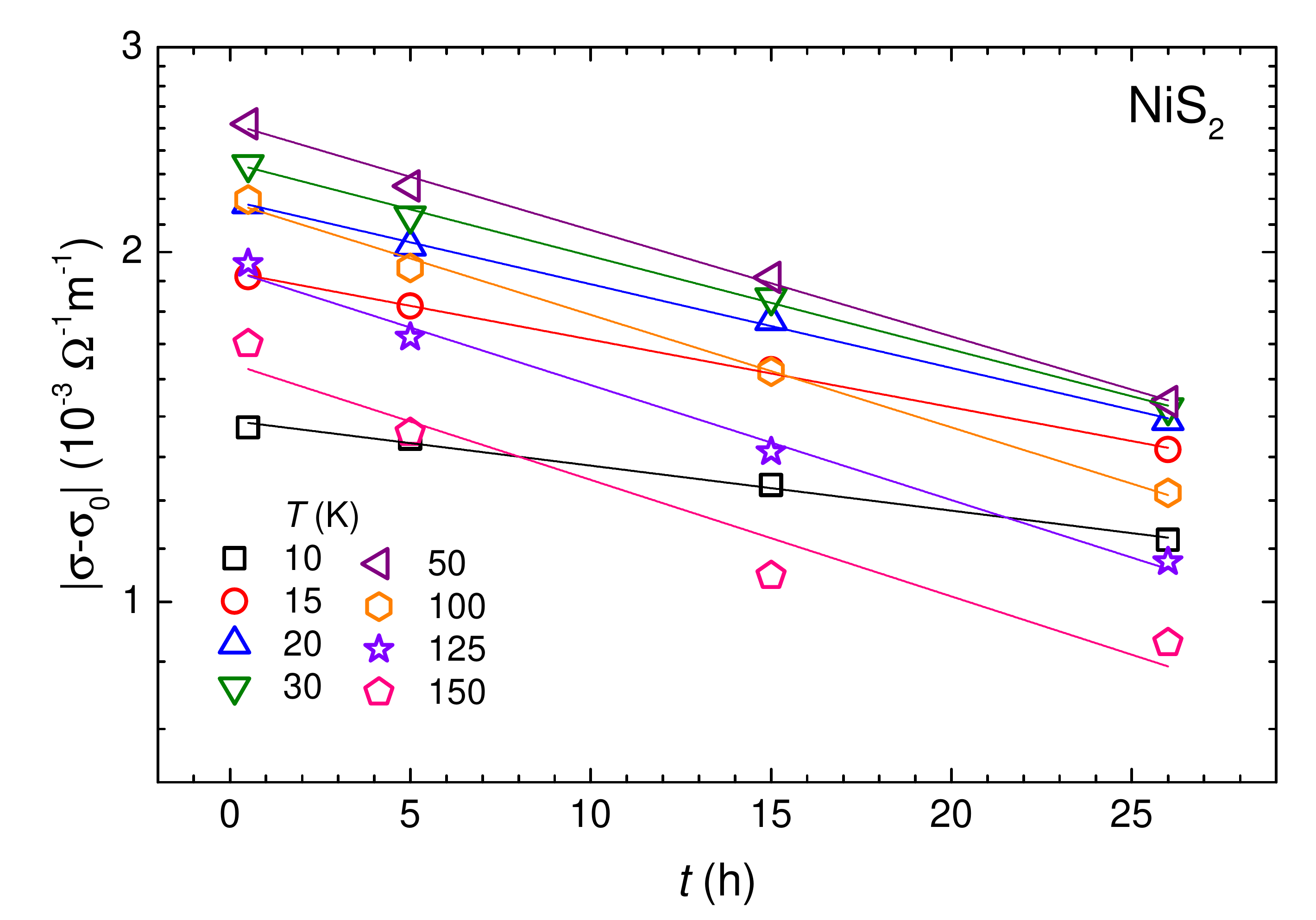}%
\caption{Relaxation of the conductivity: The difference $\left|\sigma - \sigma_0\right|$ of the conductivity with respect to the initial conductivity $\sigma_0$ is plotted against the time $t$ the sample spend at room temperature after heat treatment. Solid lines are linear fits with intercept $a$ and slope $b$ to the data in the logarithmic representation of the conductivity difference.}%
\label{fig:s0s}%
\end{figure}

The exponential nature of the decay and gap are best analysed from a logarithmic representation 
\begin{equation}
  \ln \left|\sigma-\sigma_0\right| = \ln\left(\frac{e^2 \nV}{m \nu}\right) - \frac{\Delta}{\kB T} - \frac{t}{\tau}
\label{eq:lnS}
\end{equation}
Such a logarithmic (base 10) representation is shown in Fig.~\ref{fig:s0s} where the initial conductivity $\sigma_0$ and time dependent conductivity $\sigma$ is obtained as the inverse of the resistivity of the pristine sample and after time $t$ the sample spent at room temperature subsequent to heat treatment, respectively.

\begin{figure}%
\includegraphics[width=\columnwidth]{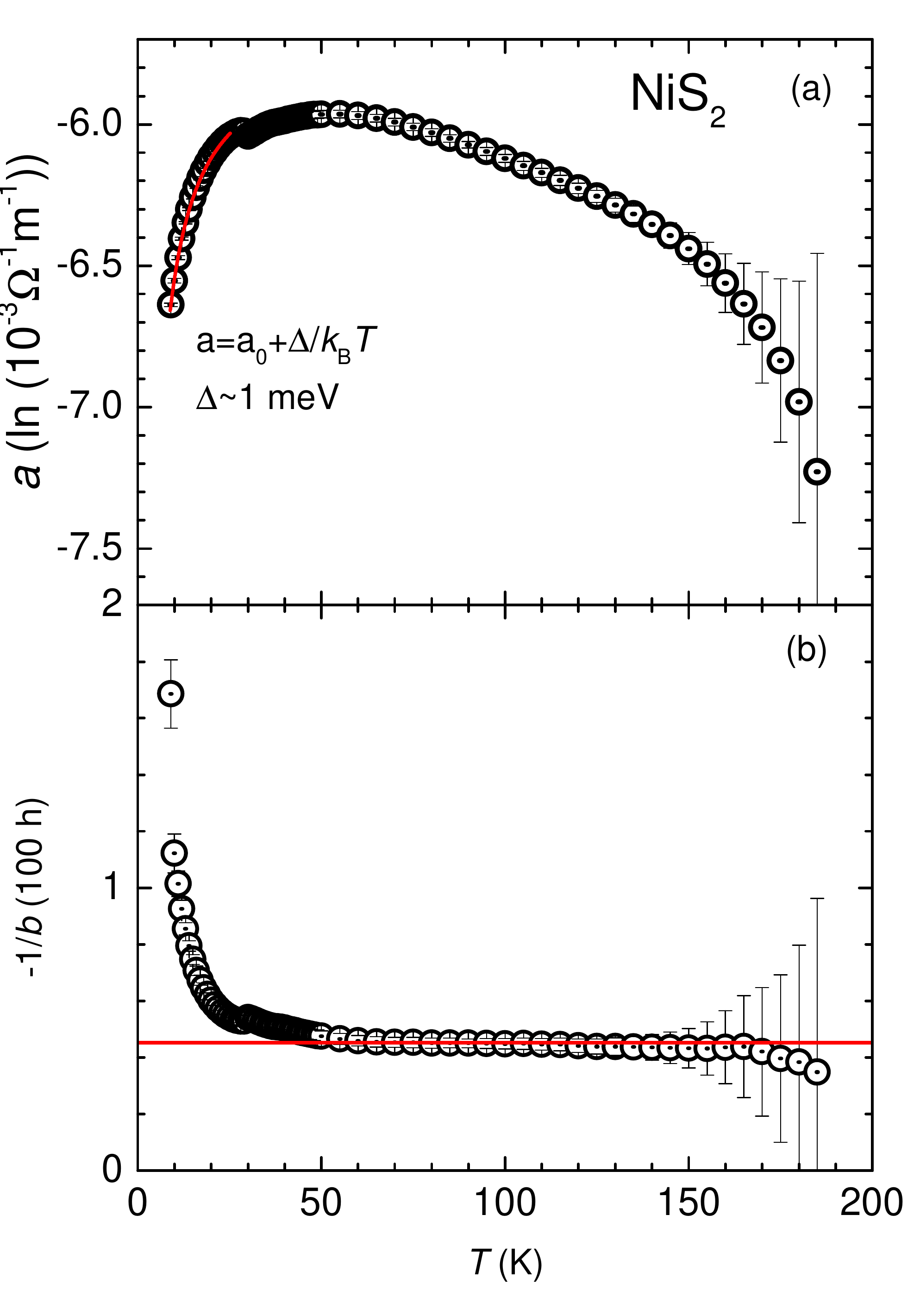}%
\caption{Temperature dependence of the intercept $a$ and inverse (negative) slope $-1/b$ from fits of $\ln\left|\sigma-\sigma_0\right|$ as shown in Fig.~\ref{fig:s0s}. Solid red line in (a) represents a reciprocal divergence $a(T)=a_0-(\Delta/\kB T)$ between \SIlist{9;20}{\K}. Solid line in (b) marks the constant relaxation time observed above \SI{50}{\K}. }%
\label{fig:FitPar}%
\end{figure}

We find the data to be well described by this logarithmic form as shown with solid lines in Fig.~\ref{fig:s0s} in agreement with our model of relaxing impurity concentration. 
The intercept $a$ and slope $b$ in Fig.~\ref{fig:s0s} represent the first two and last term in eq.~\ref{eq:lnS}, respectively:
\begin{align}
	a &= \ln\left(\frac{e^2 \nV}{m \nu}\right) - \frac{\Delta}{\kB T} & \text{intercept} \label{eq:a}\\
	b &= -\frac{1}{\tau}&\text{slope} \label{eq:b}
\end{align}
We plot $a$ and $-1/b$ in Fig.~\ref{fig:FitPar} as a function of temperature. Results in Fig .~\ref{fig:FitPar} have been converted to a natural base $e$ in line with the natural logarithmic form in eq.~\ref{eq:lnS}. Thus they represent $a$ and $-1/b$ as associated in eqs.~\ref{eq:a} and \ref{eq:b}.

The temperature dependence of $a$ and $b$ in Fig.~\ref{fig:FitPar} (a) and (b) display distinct temperature dependencies that can be linked to the behaviour of $\left|\sigma_0-\sigma\right|$ in Fig.~\ref{fig:s0s} and $\rho(T,t)$ in Fig.~\ref{fig:RvsT}: The intercept $a$ reflecting the difference $\left|\sigma_0-\sigma\right|$ immediately after heat treatment has large negative values at lowest temperature corresponding to a large difference in $\sigma-\sigma_0$ and conversely the large enhancement of the resistivity $\rho-\rho_0$ as seen in Fig.~\ref{fig:RvsT}. The intercept $a$ increases reaching a maximum at \SI{50}{\K} above which it decreases again. These two regimes are naturally associated with the two regimes of conduction in the surface: insulating below \SI{50}{\K} and metallic above \SI{50}{\K}.

Below \SI{50}{\K} in the insulating regime the second term in eq.~\ref{eq:a} is dominant governing the trapping of charge carriers. In fact, we find $a(T)$ to be compatible with a negative inverse temperature dependence as derived for the second term in eq.~\ref{eq:a}. A fit of an inverse form yields $\Delta\approx\SI{1}{\meV}$ as shown in Fig.~\ref{fig:FitPar}(a). This value agrees with the activation energy \DS\ found for the surface conduction in the insulating regime within our accuracy. Thus, this points towards scenario ii) being realised where $\Delta = \DS$.

The value of $a$ reaches a maximum around \SI{50}{\K} and decreases at higher temperatures. Whilst a non-monotonic behaviour can be understood from a temperature dependent scattering rate $\nu$ entering eq.~\ref{eq:a} the main strong downturn above \SI{100}{\K} is a signature of the diminishing surface conductivity in the metallic regime. In addition, the total conductivity is dominated by the bulk at higher temperatures resulting in the enhanced uncertainty above \SI{150}{\K}.

We now turn to the slope $b$ representing the inverse relaxation time as extracted from  fits to the relaxation of the conductivity. The relaxation time is expected to be independent of temperature in eq.~\ref{eq:b}. Indeed a constant relaxation time of $\tau=\SI{45}{\hour}$ is found above \SI{50}{\K}. Below \SI{50}{\K} we find an increase towards lower temperature which cannot be understood within our simple model. This is the regime where the surface contribution shows insulating behaviour. A variable range hopping model may be necessary to describe the behaviour in this range. Here, the time dependent impurity concentration alters the hopping distance and possibly also the energy spectrum of the hopping sites. Additionally, the presence of strong Coulomb correlations in \NiS\ may need to be taken into account as demonstrated by the unconventional exponent $x$ in the activated low-temperature regime.

\section{Discussion}
We model the relaxing behaviour of the resistivity with a reduced charge carrier concentration due to vacancies created by the heat treatment in the surface layer. Pyrites like \NiS\ are well known to generate SO$_2$ when heated (firing pyrites is an established industrial procedure to produce SO$_2$). Thus, sulphur vacancies are the most likely impurities created by heat treatment. 

We find the activation energy $\Delta$ for the missing charge carriers associated with the vacancy concentration \nV\ in eq.~\ref{eq:s0s} to be the same as the activation energy $\DS$ for surface conduction at lowest temperatures. This points towards a low level of vacancies to be present in pristine samples of \NiS\ which could arise during the growth or by storing crystals at room temperature. In this case measurements of low-temperature cleaved samples could allow detection of surface conducting in the absence of vacancies on a stoichiometric sample. 

The surface conduction is dramatically reduced at lowest temperatures. From this we conclude that the impurity concentration must be of similar magnitude as the surface carrier concentration which was estimated to be of the order of one per unit cell of \NiS\ on the surface. Assuming an impurity concentration of $\nV \approx \SI{1e14}{\per\centi\meter\squared}$ in the surface layer of unit cells we can estimate the diffusion coefficient for the sulphur to diffuse from the bulk to the surface, where we assume sulphur to diffuse from $\approx \SI{10}{unit cells}$ deep. We find a diffusion coefficient of $D\approx\SI{1e-10}{\meter\squared\per\second}$ which is large but not unprecedented with a similar diffusion coefficient of \SI{8e-11}{\meter\squared\per\second} for Cu in Si at room temperature. An Arrehnius analysis of the diffusion coefficient as a function of temperature the sample is kept at can reveal the activation energy and the mobility of the sulphur atoms. \sven{This Arrehnius analysis may be carried out in the protected atmosphere of the cryostat and below room temperature in order to avoid additional sulphur depletion.}

\section{Conclusion}
In summary, we have discovered an enhancement in the resistivity of \NiS\ induced by heat treatment. This enhancement is found to decay with the resistivity relaxing back towards the pristine behaviour with a time constant of $\tau=\SI{45}{\hour}$ at room temperature. We identify the changes to be restricted to the surface conduction of \NiS\ present at low temperatures. Our analysis suggests sulphur vacancies forming due to oxidisation of sulphur during heat treatment. These vacancies are refilled from the bulk through atomic diffusion. We estimate a diffusion coefficient $D \approx \SI{1e-10}{\meter\squared\per\second}$. This diffusion is expected to follow Arrehnius behaviour which can be tested via the dependence on temperature at which the sample is stored between low-temperature measurements. 

Our results point towards sulphur atoms to provide surface charge carriers in \NiS\ as we find the same activation energy for the pristine sample at low temperatures and for the restored charge carriers after sulphur diffusion. This suggests that activated behaviour of the surface conductivity at lowest temperatures is associated with sulphur vacancies present on the surface of pristine \NiS. 

The time dependent vacancy concentration on the surface provides a route to study variable range hopping on the surface of a correlated insulator where the length scale of the hopping can be tuned by heat treatment  on a single sample. 

In addition, our results are relevant to understand and design the properties of nano-crystalline \NiS\ and pyrite materials currently explored for applications in capacitors and batteries \cite{Pang2014,Wang2010b,Caban-Acevedo2012,Linganiso2013,Li2014a}

%
%
%
%
%
\vspace{\baselineskip}
\noindent{\bf Author information and Data management}
Correspondence and requests for materials should be addressed to S.F. (Sven.Friedemann@bristol.ac.uk). 
Raw data will be made available from the Data repository at the University of Bristol \cite{Clark2015}.

%
%
\vspace{\baselineskip}

\noindent{\bf Acknowledgements}
The authors would like to thank Chris Bell for fruitful discussions. 
We acknowledge partial support by the EPSRC.

\appendix

\section{Estimation of oxygen present during heat treatment}
\label{sec:oxygen}

We performed the heat treatment of \NiS\ in high purity atmosphere in an effort to reduce the effect of oxidisation. Yet, we estimate the amount of residual oxygen to be sufficient for full oxidisation of sulphur on the \emph{surface} of \NiS. 
After initial flushing the furnace we used a flow of $Q\approx \SI{0.5}{\milli\liter\per\second}$.  This amounts to a total volume of $\approx\SI{4}{\litre}$ passed via the sample during the heat treatment. We used \SI{99.998}{\percent} pure argon. Assuming the residual impurity to be oxygen we obtain an upper limit for the amount of oxygen present to the sample of $N_{\text O_2} \approx \SI{4e-6}{\mole}$. The sample has a surface area of less then \SI{1}{\milli\meter\squared}. This is equivalent to $\approx\SI{1e-11}{\mole}$ of \NiS\ on the surface. This amount of sulphur is 5 orders of magnitude smaller than the amount of oxygen present in the used gas. Thus, we conclude that oxygen levels are more than sufficient to cater for a surface reaction in our sample.

%


%
%



\end{document}